\documentclass{nrc1}
\usepackage{graphicx}
\usepackage{amsmath}
\usepackage{amssymb}
\usepackage[]{natbib}

\newcommand{\kms}{\ensuremath{\mathrm{km}\,\mathrm{s}^{-1}}}

\newcommand{\LCDM}{$\Lambda$CDM}

\begin{document}

\title{A Tale of Two Paradigms: \\
the Mutual Incommensurability of \LCDM\ and MOND}

\author{Stacy S. McGaugh}

\address{Department of Astronomy, Case Western Reserve University, Cleveland, OH 44106}

\shortauthor{McGaugh}

\maketitle

\begin{abstract}
The concordance model of cosmology, \LCDM, provides a satisfactory description of the evolution of the universe and the growth of large scale structure.  
Despite considerable effort, this model does not at present provide a satisfactory description of small scale structure and the dynamics of bound 
objects like individual galaxies.  In contrast, MOND provides a unique and predictively successful description of galaxy dynamics, but is mute on 
the subject of cosmology.  Here I briefly review these contradictory world views, emphasizing the wealth of distinct, interlocking lines of evidence that went into
the development of \LCDM\ while highlighting the practical impossibility that it can provide a satisfactory explanation of the
observed MOND phenomenology in galaxy dynamics.  I also briefly review the baryon budget in groups and clusters of galaxies where neither paradigm
provides an entirely satisfactory description of the data.  Relatively little effort has been devoted to the formation of
structure in MOND; I review some of what has been done.  While it is impossible to predict the power spectrum of the microwave background temperature
fluctuations in the absence of a complete relativistic theory, the amplitude ratio of the first to second peak was correctly predicted a priori.  
However, the simple model which makes this predictions does not explain the observed amplitude of the third and subsequent peaks.  
MOND anticipates that structure forms more quickly than in \LCDM.  This motivated the prediction that reionization would happen earlier in MOND 
than originally expected in \LCDM, as subsequently observed.  This also provides a natural explanation for 
massive, early clusters of galaxies and large, empty voids.  However, it is far from obvious that the mass spectrum
of galaxy clusters or the power spectrum of galaxies can be explained in MOND, two things that \LCDM\ does well.  
Critical outstanding issues are the development of an acceptable relativistic parent theory for MOND, and the reality of the 
non-baryonic dark matter of \LCDM.  Do suitable dark matter particles exist, or are they a modern \ae ther?
\end{abstract}


\begin{quote}
\textit{A single galaxy might seem a little thing to those who consider only the immeasurable vastness of the universe, 
and not the minute precision to which all things therein are shaped.} --- paraphrased from the Ainulindal\"e by J.R.R.\ Tolkien
\end{quote}

\section{Introduction}

There is copious evidence for mass discrepancies in the universe \cite{sandersbook}.
The usual dynamical laws --- specifically, Newton's law of gravity and Einstein's generalization thereof --- 
fail when applied to the visible mass in galaxies, clusters of galaxies, and the universe as a whole.
These are well tested, fundamental theories, so the observed discrepancies are usually attributed to unseen mass.

The evidence for dark matter remains entirely astronomical in nature. 
The existence of dark matter is an inference based on the reasonable
assumption that General Relativity suffices to describe the dynamics of the universe and its contents.
The same evidence calls this assumption into question.

\subsection{What's in a name?  The Missing Mass problem or the Acceleration Discrepancy?}

The need for dark matter is often referred to as the `missing mass problem.' 
This terminology prejudices the answer.  More appropriately, we might call it the mass discrepancy problem: there is 
either missing mass, or a discrepancy in the equations that lead to its inference \cite{milgrom83}.
It has also been suggested \cite{bekensteinreview} that the proper terminology should be the acceleration discrepancy,
as the problem manifests at a nearly universal acceleration scale \cite{LivRev}.

Though General Relativity is well tested in a variety of precise ways, the only data that test it on the scales of galaxies
are the data that show the discrepancy.  If we drop the assumption that General Relativity applies
in regimes where it is otherwise untested, much of the evidence for dark matter becomes rather ambiguous.
Indeed, the data can often be interpreted as well in terms of a modification of dynamics as dark matter \cite{MdB98a,MdB98b}.
Some data favor one interpretation and some the other, and which we choose is dictated entirely by how we weigh
disparate lines of evidence.

\subsection{Philosophical Aspects}

The current situation is reminiscent of that described by Kuhn \cite{kuhn}.
Proponents of competing para\-digms have different ideas about the importance of solving different scientific problems, 
and about the standards that a solution should satisfy.
The vocabulary and methods of the paradigms differ to the point where they can become mutually incomprehensible.

The concordance cosmology is about the geometry and expansion history of the universe.
The language is that of metrics and power spectra.  
From this perspective, galaxies are small things that serve best as tracers of the large scale structure that extends across 
the immeasurable vastness of the universe.

In contrast, MOND \cite{milgrom83} is about the dynamics of objects in the low acceleration regime.  
The language is that of gravitational potentials and orbital mechanics.
From this perspective, the dynamics of individual galaxies are connected to their observed shapes with minute precision.

The contradictory concepts of dark matter and modified gravity have been developed into paradigms that enjoy successes and suffer
problems in different arenas.  Presumably one must subsume the other, but it is far from obvious how this can be achieved short of
a hybrid solution \cite{sandersbosons,blanchet,Blanchetnatural,angussterile}, which some might find aesthetically distasteful.
To illustrate the situation, I briefly state the impossibility of each alternative from the perspective of the other.

\section{The solution must be dark matter}
\label{sec:DMright}

\begin{figure}
\includegraphics[width=5.5in]{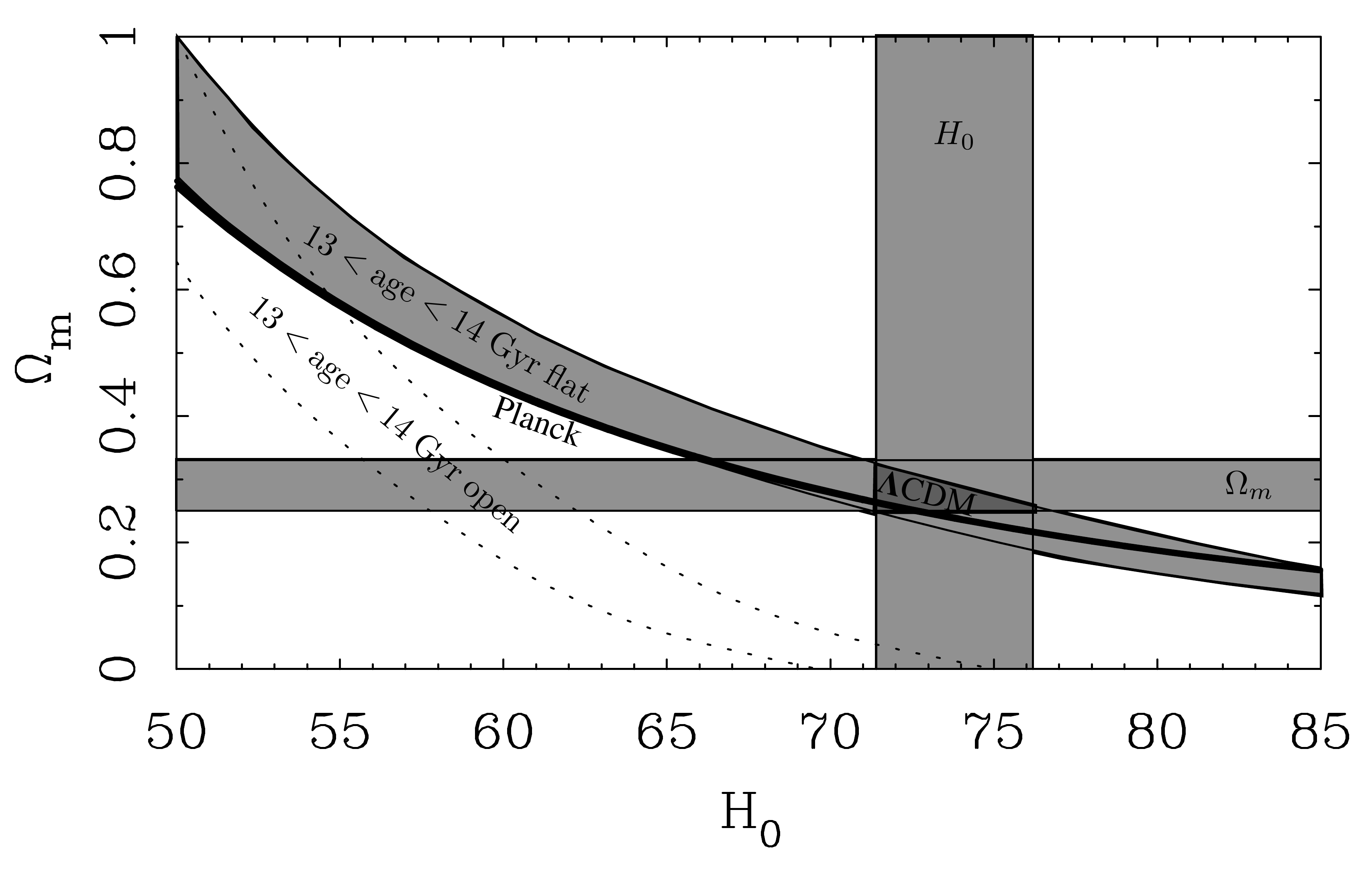}
\caption{The cosmic mass density $\Omega_m$ and expansion rate $H_0$ (in $\kms\,\mathrm{Mpc}^{-1}$).  
Constraints on these parameters are shown as gray bands.  Both the expansion rate and the matter density have been measured directly
many times \cite{HSTKP,riessH0,OmegaDavis}.  Also shown is the range expected from the ages of the oldest stars, 13 -- 14 Gyr \cite{oldeststars}.  
The dotted lines show this expectation for an open, low density universe, while the shaded band shows it for a flat universe in 
which $\Omega_m + \Omega_{\Lambda} = 1$.  Consistency between all three requires a cosmological constant, and constrains
cosmology to the small region labeled \LCDM\ (dark gray triangle).  This was known by the mid-90s \cite{OS}; more recently, the Planck satellite
constrains cosmology to lie within the black band $\Omega_m h^3 = 0.0959 \pm 0.0006$ \cite{planckXVI}.  
Note that the width of the band illustrates the allowed range: parameter combinations outside this band are now disallowed.}
\label{OmH}
\end{figure}

The current cosmological paradigm, \LCDM, is the culmination of a wealth of interlocking lines of evidence that all point to a single solution.
The universe is clearly expanding, with the rate of expansion $H_0$ being well measured thanks in large part to the Hubble space telescope \cite{HSTKP,riessH0}.
The mass density $\Omega_m = \rho/\rho_{crit}$ 
of the universe has persistently been estimated to lie in the range $1/4 < \Omega_m < 1/3$ \cite{OmegaDavis,White93}.
These two fundamental parameters of cosmology are shown in Fig.~\ref{OmH}.
 
The age of the oldest stars is approximately 13 to 14 Gyr \cite{oldeststars}.  This can be plotted as a further constraint in Fig.~\ref{OmH}.  
A consistent picture emerges, provided that we compute the age of a geometrically flat universe rather than an open one: 
$H_0 = 72\;\kms$ and $\Omega_m \approx 1/4$.  These constraints were already well established by the mid-1990s.  
Several other lines of evidence also pointed in the same direction at that time \cite{OS}, including cluster baryon fractions \cite{White93},
faint galaxy counts \cite{yoshii}, the early appearance of structure \cite{MoFukugita}, and the galaxy power spectrum \cite{karlfisher}.

In order to have a geometrically flat universe with $\Omega_m < 1$, it is necessary to invoke the cosmological constant $\Lambda$ such that
$\Omega_m + \Omega_{\Lambda} = 1$.  This concordance model of cosmology, \LCDM, makes a strange and unlikely prediction. 
In effect, $\Lambda$ behaves like anti-gravity, so the expansion rate of the universe should be accelerating \cite{negativeq}.  
This was subsequently inferred to be true from observations of Type Ia supernova \cite{riess98,perlmutter}.
Further confirmatory evidence was provided by observations of temperature fluctuations in the cosmic microwave background (CMB) which 
showed that the geometry is indeed very nearly flat \cite{boomerang,WMAP7} in the sense of the Robertson-Walker metric.  
The inferred value of $\Lambda$ is in good accord with the other observational constraints, which have now become very precise (Fig.~\ref{OmH}).  
Consequently, the concordance model seems irrefutable.

An essential component of the concordance model is non-baryonic cold dark matter (CDM).  This form of dark matter must be composed of 
some novel non-baryonic material.  It is usually assumed to be some Weakly Interacting Massive Particle (WIMP), 
but the specific candidate particle is not important here.  What does matter is that there is
a form of mass that outweighs normal matter by roughly a factor of 5 \cite{WMAP7}, does not interact electromagnetically nor participate in
Big Bang Nucleosynthesis (BBN), and is dynamically cold in order to seed the formation of large scale structure.  These requirements provide
compelling reasons why the dark matter must be non-baryonic.

\begin{figure}
 \includegraphics[width=5.5in]{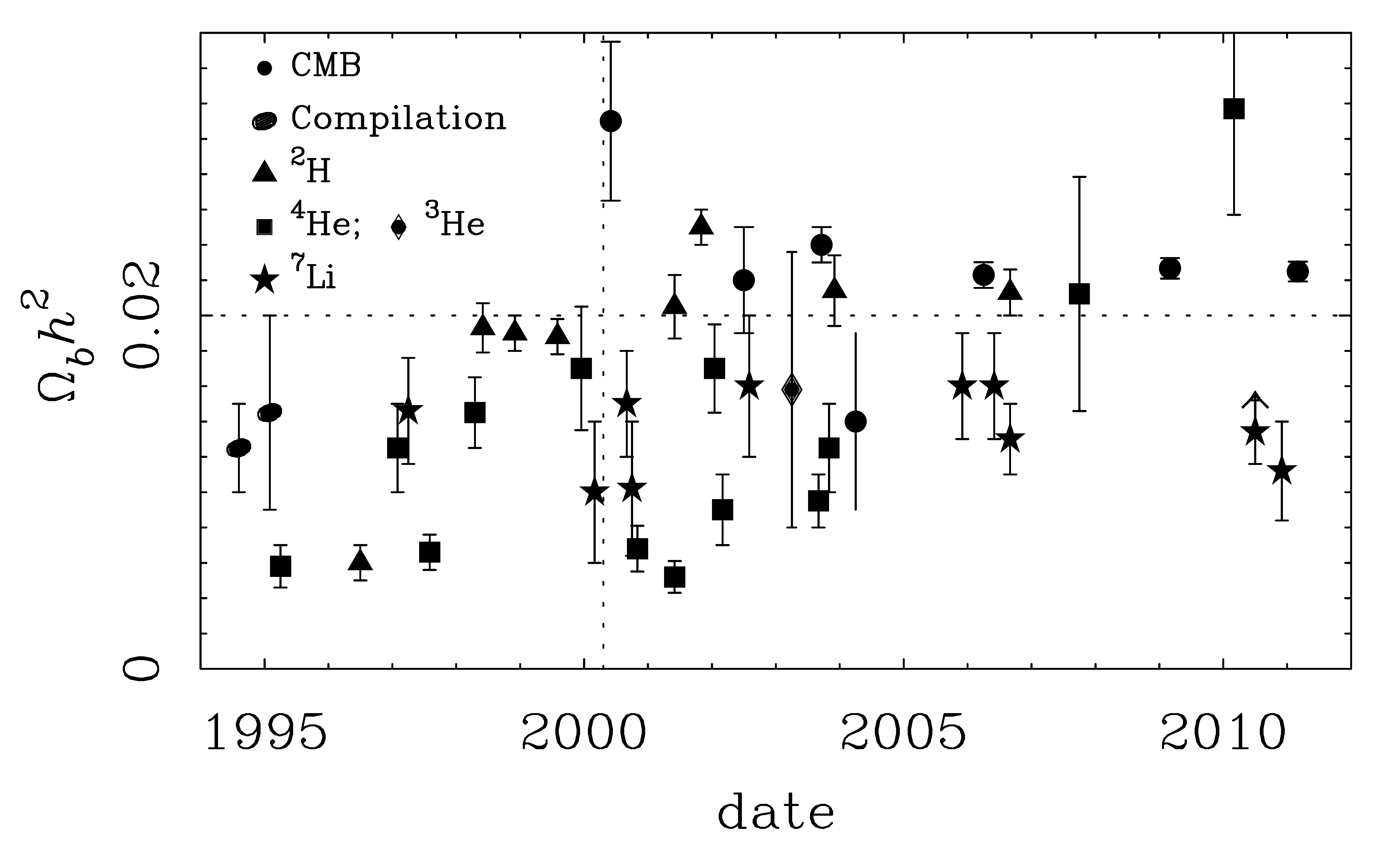}
  \caption{The baryon density from various measurements over the past two decades, as tabulated by \cite{M04}, with a few more 
  recent updates \cite{WMAP7, CC, asplund, boni, newD, PLP}.  
  Constraints from each independent method are noted by different symbols, as noted in the inset.  
  The data paint a broadly consistent picture --- BBN is almost certainly correct --- but important tensions remain, most notably 
  between lithium and post-CMB deuterium \cite{cyburtlithium}.
  BBN was well established long before the start date of this graph; previous work is represented by early compilations.  
  The first point is the famous $\Omega_b h^2 =0.0125$ value \cite{walkerBBN}.  
  Note that no measurement of any isotope ever suggested a value $\Omega_b h^2 > 0.02$ (horizontal dotted line) 
  prior to the appearance of relevant CMB data (vertical dotted line).  No fit to the CMB that includes CDM tolerates $\Omega_b h^2 < 0.02$. 
  Measured values of some isotopes seem to have drifted upwards towards the CMB values since 2000, a worrisome trend given the 
  dangers of confirmation bias.  Lithium (derived from stellar rather than cosmic observations) has not followed this trend.}
  \label{fig:BBN}
\end{figure} 

The formation of the light elements in the first few minutes of the expansion of the universe is one of the empirical pillars of the Big Bang.
Observations of the primordial abundances of the isotopes of hydrogen, helium, and lithium provide a strong constraint on the baryon-to-photon ratio.
Since the photon energy density is well measured by the temperature of the CMB, these observations effectively constrain the baryon density, 
$\Omega_b h^2$ (where $h = H_0/100\;\kms\,\mathrm{Mpc}^{-1}$). Fig.~\ref{fig:BBN} shows the baryon density estimated from many different methods.  
It is persistently low: $\Omega_b \approx 0.05$.  In contrast, the total gravitating mass density is $\Omega_m \approx 0.25$.  
Since $\Omega_m > \Omega_b$, there is a lot of mass that gravitates but which is not baryonic.

Structure formation similarly requires some form of non-baryonic gravitating mass.  The initial condition written on the CMB is that of a hot plasma that is initially
homogeneous to one part in $10^5$ \cite{WMAP7,planckXVI}.  From this very modest starting point, the universe grew the vast diversity of large scale structure,
from individual galaxies and their contents to rich clusters, long filaments, and enormous voids \cite{LSSbook,CfAredshift,shethLSS}.  Though the enormity of these 
structures initially came as a surprise, they are the natural result of the gravitational growth of initial density fluctuations.  The challenge is the amplitude:
baryons are coupled to the photon field at early times, and can grow at most $\sim 10^3$ in a Hubble time.  What is needed is a form of mass that does
not couple to the photons so it can begin to form structure sooner, and already have a large amplitude at the time of recombination without leaving too
obvious a mark on the CMB.  A dose of non-baryonic and dynamically cold dark matter fits the bill nicely.

Both BBN and structure formation provide strong arguments in favor of non-baryonic dark matter.  Together, these arguments were sufficiently compelling to convince me
(and most of the community) that CDM had to exist.  From a cosmological perspective, the ability of particle physics models to provide candidates for the CDM
is merely icing on the cake:  the need for such stuff is clear, so surely some candidate\footnote{That supersymmetry increasingly looks like a good idea that Nature 
declined to implement is perhaps more worrying from a particle physics perspective.} (WIMP or other) will win out.  Flashing forward a quarter century, CDM does leave 
a definite (if subtle) mark on the CMB.  Formal fits obtain $\Omega_{\mathrm{CDM}} > 0$ at $44\sigma$ \cite{planckXVI}.
Given the many compelling successes of the concordance cosmology, the various interlocking but distinct lines of evidence supporting it,
and its basis in the highly successful theory of General Relativity, it seems quite unreasonable to doubt the existence of CDM.

\section{The solution must be modified dynamics}
\label{sec:MONDright}

The two arguments that launched the CDM paradigm, $\Omega_m > \Omega_b$ and the growth of structure by a factor of $10^5$ 
rather than $10^3$ are both persuasive and well established.  However, they only hold under the imminently reasonable assumption that gravity is normal.  
If we drop this assumption, neither is conclusive.

If a modification of gravity is responsible for the observed mass discrepancies in the universe, 
it is easy to imagine that a conventional measurement of the mass density is
an overestimate that is an artifact of applying the wrong force law. In MOND \cite{milgrom83}, the dynamically estimated $\Omega_m \approx \Omega_b$ \cite{MdB98b}.  
Similarly, changing the force law can result in the faster growth of structure \cite{sanderscosmo,MearlyU}.
The standard analysis of the CMB power spectrum \textit{assumes} General Relativity plus CDM, so it would be circular to interpret 
the resulting $\Omega_{\mathrm{CDM}} > 0$ as a detection of dark matter.  
All this really confirms is that General Relativity \textit{cannot} describe the observed universe with only normal baryonic matter.

Empirically, the mass discrepancy in large, gravitationally bound objects is manifested by an excess of speed \cite{sandersbook}.  
Spiral galaxies rotate faster than can be explained by the observed stars and gas.  
The stars in tiny dwarf galaxies move faster than can be contained by the gravity of the stars alone.  
Similarly, individual galaxies in rich clusters of galaxies move so fast that the clusters should fly apart in a fraction of a Hubble time. 
These objects only remain bound if there is some extra force above and beyond that which is obvious.

\begin{figure}
\includegraphics[width=5.5in]{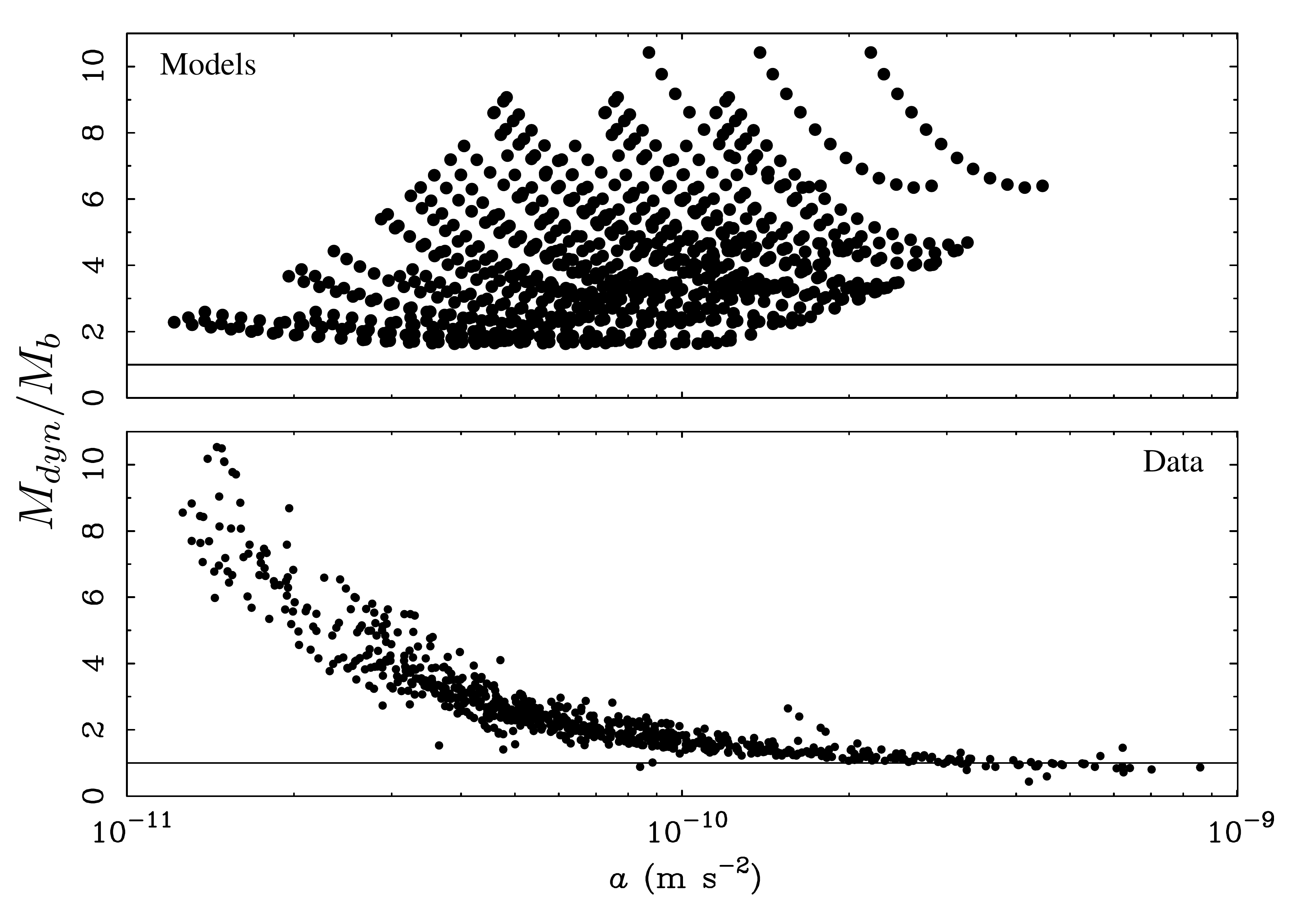}
\caption{The mass discrepancy--acceleration relation.  
The ratio of dynamical to baryonic mass is shown at each point along rotation curves as a function of the centripetal acceleration at that point.  
The top panel shows model galaxies in \LCDM\ (see text).  The bottom panel shows data for real galaxies \cite{MDacc}.  
Individual galaxies, of which there are 74 here, do not distinguish themselves in this diagram, though model galaxies clearly do.
The organization of the data suggest the action of a single effective force law in disk galaxies.  
This phenomenon does not emerge naturally from \LCDM\ models.}
\label{fig:MDacc}
\end{figure}

The signature of modified gravity in the data is the appearance of the mass discrepancy at a particular physical scale.  
This should not happen in CDM, which is scale free.  Yet such a scale is observed in the data (Fig.~\ref{fig:MDacc}).  

The relevant scale is not one of size, as one might initially suspect.  The critical scale turns on to be one of acceleration 
(or equivalently, surface density, since $a \propto G \Sigma$).
Above the critical acceleration, of order $a_{\dagger} \approx 10^{-10}\;\mathrm{m}\,\mathrm{s}^{-2}$ \cite{BBS,M11}, 
there is essentially no evidence for dark matter.  Below this scale, the mass discrepancy appears.  The severity of the mass discrepancy --- 
effectively just the ratio of total dynamical mass to observed baryonic mass --- increases systematically with declining acceleration.

A curious consequence of the mass discrepancy--acceleration relation is that the dynamics of a galaxy can be predicted simply by looking at it.
The baryonic surface density $\Sigma_b$ is predictive of the amount of dark matter required.  The lower the surface density of stars, the lower the
acceleration, and the more dark matter is implicated.  
This leads to the paradoxical situation that even in low surface brightness galaxies, which are inferred to be dark matter
dominated throughout \cite{MdB98b}, no knowledge of the dark matter is necessary to predict the dynamics it supposedly dominates.  
Observation of $\Sigma_b$ suffices.

The mass discrepancy--acceleration relation is observationally well established and purely empirical \cite{MDacc}.
It implies that there is a single effective force law in disk galaxies.
The natural interpretation is that this is indeed caused by a universal force law.
The one that works happened to already have a name: MOND \cite{milgrom83}.

The observed mass discrepancy--acceleration relation does not occur naturally in \LCDM.
Indeed, \LCDM\ makes no clear prediction for individual galaxies.  One must resort to model building.
The argument then comes down to what constitutes a plausible model.
I have spent many years trying to construct plausible \LCDM\ models.  I have never published any, because none are satisfactory.
All I can tell you so far is what does not work.

The first model that did not work was SCDM.  In the mid-1990s, the Standard CDM cosmological model had $\Omega_m = 1$ and no cosmological constant.
The amplitude of the rotation curve predicted by this model \cite{NFW} was far too high to explain observed galaxies.  This can be alleviated by lowering $\Omega_m$
($\Lambda$ is largely irrelevant to this problem).  Consequently, \LCDM\ is closer to the right ballpark, though it still over-predicts velocities \cite{KdN08,KdN09}.
Formally, one needs $\Omega_m < 0.2$ \cite{MBdB03}, which is unacceptable to CMB data \cite{WMAP7,planckXVI}.  This problem motivates an
entire sub-field of work on feedback processes in galaxy formation that seek to redistribute mass in order to dodge this constraint.

In Fig.~\ref{fig:MDacc}, I show what I consider to be a natural \LCDM\ model.  In this structure formation paradigm, dark matter perturbations are already
growing structure at the time of recombination, creating dark matter halos for the baryons to fall into once the photon field releases them from its grip.
The baryons dissipate and sink to the centers of the the non-dissipative dark matter halos where they form galaxies.  In this process, they are expected
to drag some dark matter along with them, a process referred to as adiabatic contraction.  The models shown in Fig.~\ref{fig:MDacc} are
built by adiabatically compressing \cite{adiabat} NFW halos \cite{NFW} drawn from the range expected for halos in 
structure formation simulations \cite{maccio2008} with exponential stellar disks representing a portion of the observed range of size and stellar mass \cite{dJL2000}.
I do not show the full range of observed properties because that simply increases the scatter further, making the agreement with data worse.
The model clearly fails, so is not worth pursuing.

The failure of the natural \LCDM\ galaxy formation model drives simulators to consider feedback.  Feedback in the context of galaxy formation invokes the
energy created by baryonic processes like supernovae to rearrange the distribution of mass in model galaxies.  This is an inherently chaotic process, so it
does not naturally lead to the observed organization.  There are many groups at present working on simulations of feedback in galaxy formation, but I am only
aware of one claim to demonstrate the observed organization \cite{vdBD2000}.
Such models\footnote{The model of \cite{vdBD2000} was motivated by an early version of the empirical relation \cite{myrutgers}, which was not
anticipated in \LCDM.} are of necessity highly fine-tuned \cite{MdB98a}.  

Fine-tuning is always possible in dark matter models.  There are many free parameters, and we are always free to add more.  So I do not doubt that
it is possible to mimic the data in Fig.~\ref{fig:MDacc}.  The question then becomes whether the real universe operates that way.  My fear is that feedback
has become a modern version of the epicycle:  a \textit{deus ex machina} that is invoked to excuse any failing of the standard model, no matter how bizarre.

The data show the operation of a single effective force law in galaxies.  There is no reason to expect this behavior with dark matter.  Indeed, the combination
of a thin, spiral disk of stars and gas with a quasi-spherical halo of non-baryonic CDM can in principle result in a vast variety of effective force laws.
Yet from the enormous parameter space available to \LCDM, Nature only selects the peculiar one specific to MOND.

In the solar system, the effective force law is Newton's inverse square law (modulo the tiny
excess perturbation on the precession of Mercury's perihelion caused by General Relativity).  Everywhere we probe in the solar system\footnote{Note that
the solar system is well above the critical acceleration scale.  For example, the Earth orbits with a centripetal acceleration of $5 \times 10^7\;a_{\dagger}$.}, 
the inverse square law works perfectly.  If someone were to suggest that really the solar system operates on an inverse cube law, and it merely looks like
an inverse square law because there exists dark matter that is always arranged to make it look just so, would we consider this suggestion reasonable?
This is the situation in galaxies, where we observe a single effective force law but attribute it to dark matter that happens to be arranged just so.

MOND has had many \textit{a priori} predictions come true.  It is the only theory to anticipate correctly in advance the dynamical behavior of low surface brightness
galaxies \cite{MdB98b,dBM98}. It makes specific predictions for the detailed mass distribution and velocity structure of the Milky Way \cite{M08,BFWZA} that have 
subsequently been realized \cite{bovyrix,binneyrave}.  It makes its strongest predictions deep in the modified regime of low accelerations.  
These are well probed by nearby dwarf spheroidal galaxies, for which it was possible not only to explain the observed velocity dispersions based on the 
observed distribution of mass \cite{GS92,milg7dw,angusdw,serradw,MM13a,Fabianclassical}, but predict them in advance \cite{MM13b,PM14}.  
This prediction cannot even be made in the context \LCDM.
A distinctive prediction is made for tidal dwarf galaxies by both theories, where again MOND is successful \cite{gentiletidal}.
Given the obvious organization of the data, the many predictive successes of MOND,
and the inability of \LCDM\ in many cases to make comparable predictions, it seems quite unreasonable to believe in the existence of invisible mass from beyond
the Standard Model of particle physics that pervades the universe with nary a signal in the laboratory.

\section{Ambiguities in groups and clusters of galaxies}
\label{Ambiguous}

The conclusions of \S \ref{sec:DMright} and \S \ref{sec:MONDright} are diametrically opposed.
At present, it is not possible to reconcile all lines of evidence.  One is therefore left to weigh the various lines of evidence as seem best.
The answer then follows from whatever weighting scheme is chosen \cite{kuhn}.  That much is fine, but it can lead to an unhealthy attitude in which
we simply assume our most favored paradigm is correct and all else must follow.  If we are convinced that \LCDM\ is correct, then modeling galaxies is a pesky
distraction rather than a fundamental problem to address head on.  Similarly, if we are convinced that MOND is correct, then \LCDM\ is simply 
the best conventional proxy for the true cosmology of the underlying relativistic theory of MOND.  

\begin{figure}
\includegraphics[width=5.5in]{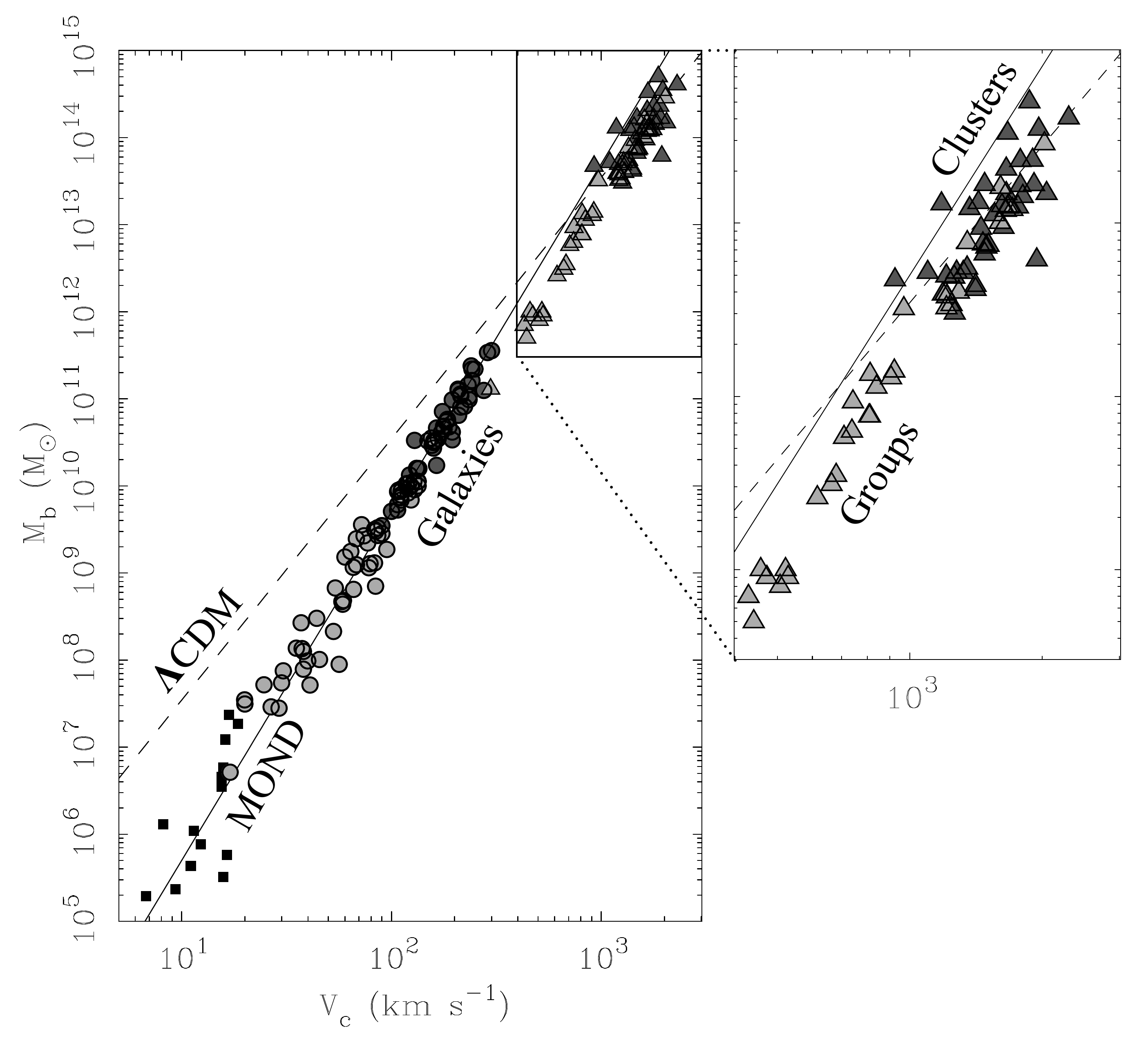}
\caption{Baryonic mass as a function of circular velocity for systems ranging from dwarf spheroidals \cite[squares]{MWolf} 
through gas rich \cite[light gray circles]{M12} and star dominated \cite[dark gray circles]{M05} spiral galaxies to groups \cite[light gray triangles]{angusbuote}
and clusters \cite[dark gray triangles]{sanders2003} of galaxies.  
The prediction of MOND is shown as a solid line and the nominal expectation of \LCDM\ is shown as a dashed line.
MOND describes the data well over six decades in mass.
The inset is expanded at right to illustrate that \LCDM\ provides a better description of the richest clusters of galaxies.  
Neither theory provides an entirely satisfactory description of groups.  The data more closely follow the 
line of constant acceleration ($M \propto V_c^4$) expected in MOND at all scales
--- nearly ten decades in baryonic mass.  
The discrepancy with the slope predicted by \LCDM\ for objects smaller than clusters of galaxies leads to the inference of missing baryons (in addition to 
non-baryonic CDM) in each and every dark matter halo \cite{M10}.}
\label{BTF}
\end{figure}

Having briefly reviewed data that support each paradigm, I discuss here some data whose interpretation is ambiguous. 
I have reviewed these many times previously \cite{MdB98b,SMmond,LivRev}, and cannot revisit all issues.
I concentrate here on clusters of galaxies, which, crudely speaking, are the scale where reality transitions from looking like MOND to looking like \LCDM.

Fig.~\ref{BTF} shows the relation between observed baryonic mass and rotation velocity.  Data for various objects are shown, ranging from 
tiny dwarf satellite galaxies through giant spiral galaxies to groups and rich clusters of galaxies.  Also shown are the predictions of MOND and \LCDM.
In the latter case, I utilize \cite{M10,M11,M12} the predicted virial mass--virial velocity relation for CDM halos \cite{NFW} with the cosmic baryon fraction $f_b = 0.17$ \cite{WMAP7}.

At the scales of the richest clusters, nearly all the baryons expected in \LCDM\ are detected.
That is, the ratio of baryonic to total mass is consistent with the cosmic value.
On smaller scales, Fig.~\ref{BTF} shows that there are many fewer detected baryons in galaxy groups and
individual galaxies than are available in their dark matter halos.  
This halo-by-halo missing baryon problem \cite{cardiff,M10} is not subtle.
It grows more severe in smaller objects, being a factor of $\sim 3$ for bright galaxies like the Milky Way, growing to a factor of $> 50$ in the smallest dwarfs.
This is generically explained as the result of feedback expelling the excess baryons, or at least preventing them from cooling to form stars \cite[see discussion in][]{M12}.

\begin{figure}
\includegraphics[width=5.5in]{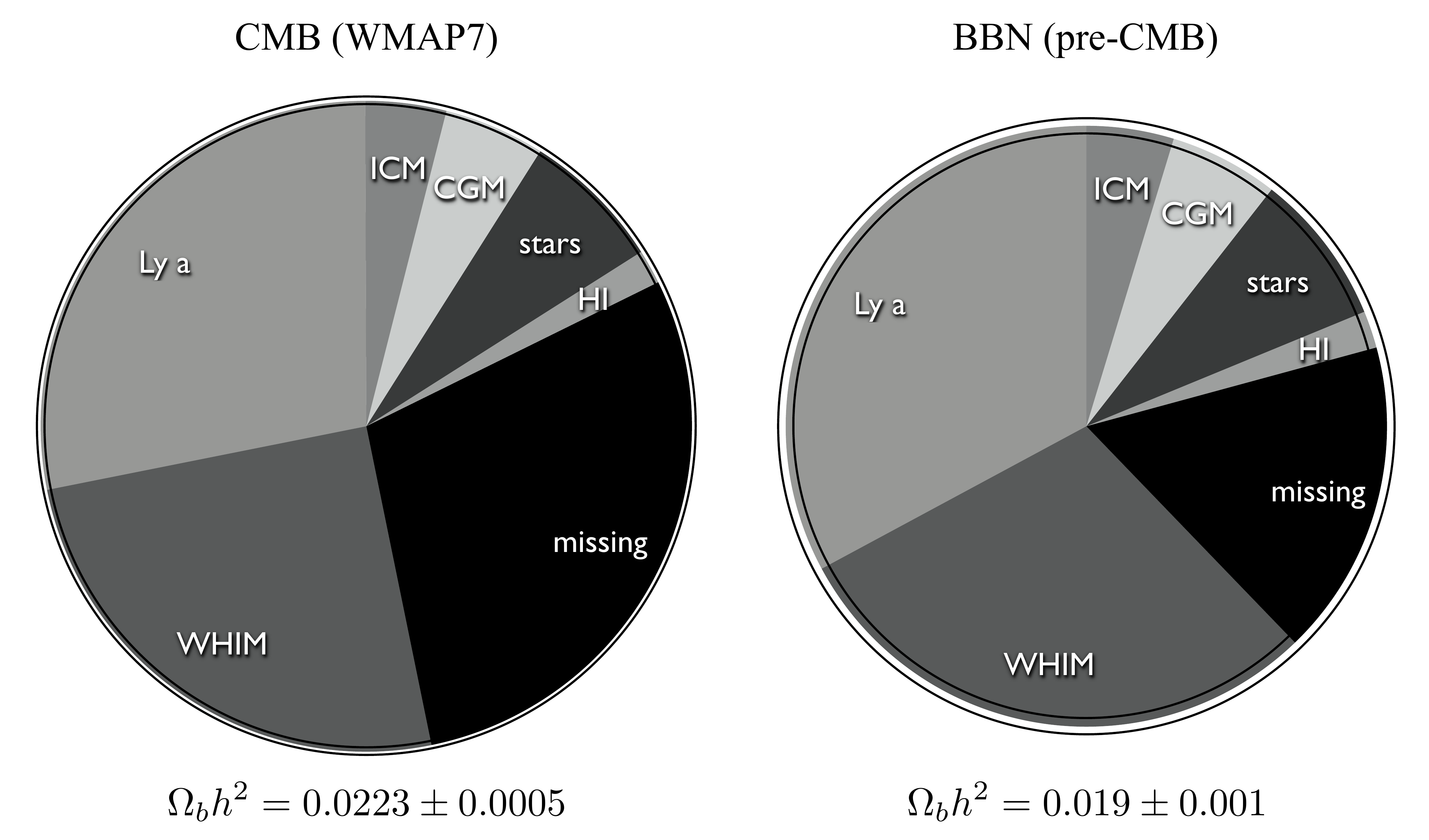}
\caption{The baryon budget in the low redshift universe adopted from \cite{missingbaryons}. The census of baryons includes the detected 
Warm-Hot Intergalactic Medium (WHIM), the Lyman $\alpha$ forest, stars in galaxies, detected cold gas in galaxies (labeled HI for atomic gas but including 
molecular $H_2$), other gas associated with the environments of field galaxies (the Circumgalactic Medium, CGM), 
and the Intracluster Medium (ICM) of groups and clusters of galaxies. 
The sum of known baryons ($\Omega_b h^2 \approx 0.016$) falls short of the density of baryons expected from BBN.  The severity of the shortfall depends on 
the total baryon density:  $\Omega_b h^2 = 0.0023 \pm 0.0005$ from fits to the CMB acoustic power spectrum \cite{WMAP7}, in which case $\sim 30\%$ of the
expected baryons are missing (pie chart on left).  For $\Omega_b h^2 = 0.019 \pm 0.01$ from BBN prior to CMB measurements \cite{tytler2000},
$\sim 17\%$ are missing (pie chart at right).  The area of the pie charts has been scaled to reflect the difference in absolute baryon abundance.  The circles near
the edge of each pie chart illustrate the formal uncertainty in the baryon density: they are, in effect, the error bars on the area.
The missing baryons presumably exist in some as yet undetected (i.e., dark) form. If a fraction of these dark baryons reside in clusters 
(in an amount a bit larger than in the ICM), it would suffice to explain the residual mass discrepancy MOND suffers in clusters of galaxies.}
\label{baryoncontent}
\end{figure}

In MOND, there is no missing baryon problem in galaxies, which behave as predicted over six decades in mass.  However, MOND suffers a missing mass
problem in groups and clusters of galaxies \cite{TheWhite,sanders2003,angusbuote}.  This is apparent in Fig.~\ref{BTF} as a shift of the data away from the 
line representing MOND.  This is a generic effect that is illustrated in dramatic fashion by the particular case of the bullet cluster \cite{clowe,angusTVS,angusgrav}.

This residual mass discrepancy in MOND is a serious problem.  
Though the discrepancy is not as large in amplitude ($\sim 2.5$) as in \LCDM, one is obliged to invoke some form of unseen mass.
This is embarrassing for an idea that stands in contrast to that of cosmic dark matter.  Indeed, the bullet cluster has been called `direct proof' of dark matter \cite{clowe}.

Here we encounter an epistemological problem.  
There is nothing about the data for clusters that informs us that the unseen mass must be \textit{the}
non-baryonic CDM required by cosmology.  Just as we are obliged accept that there are dark baryons as well as non-baryonic CDM in \LCDM,
in MOND we are also obliged to invoke dark baryons in clusters, or some more exotic form of unseen mass like heavy normal \cite{sanders2007} or 
sterile \cite{angussterile} neutrinos.

Not all baryons necessarily exist in some readily detected form.  Both theories apparently require some dark baryons.  This is also
required by BBN, which indicates the existence of more baryons than are detected directly in the local universe \cite{missingbaryons}.

An obvious question is whether BBN permits the existence of sufficient baryons to explain the residual mass discrepancy in clusters.  
Fig.~\ref{baryoncontent} shows the distribution of known baryons summarized by \cite{missingbaryons}.  Over half of all baryons reside in the intergalactic medium,
either in the WHIM or as part of the Lyman $\alpha$ forest.  Galaxies (including stars and HI gas) account for a modest slice, as does the ICM
of clusters of galaxies.

Fig.~\ref{baryoncontent} shows two pie diagrams, one for $\Omega_b h^2 = 0.0223$ as appropriate for \LCDM\ and another for 
$\Omega_b h^2 = 0.019$, which I consider more appropriate for MOND (see \S~\ref{cmbfluctuations}).  
The difference arises from fitting the CMB in the first case \cite{WMAP7}, or adopting the BBN value as it existed
prior to relevant CMB observations \cite[see Fig.~\ref{fig:BBN}]{tytler2000}.  In both cases, a large fraction of the expected baryons are missing.

The observed sum of baryons amounts to $\Omega_b h^2 \approx 0.016$ \cite{missingbaryons}.
In order to address the missing baryon problem in clusters in MOND, we need an amount that is about half again as large as the ICM wedge in Fig.~\ref{baryoncontent}.
This hardly makes a dent in the missing baryon problem if $\Omega_b h^2 = 0.0223$.  For $\Omega_b h^2 = 0.019$, it accounts for maybe half of the missing baryons.

Given the need from BBN for unseen baryons, we cannot use the cluster data to confirm the existence of CDM, nor use it to refute MOND.
Independently of BBN, both \LCDM\ and MOND require some unseen baryons, just in different places.  In \LCDM, it would be convenient for these baryons
to be associated in some way with the dark matter halos of individual galaxies, perhaps as part of the CGM.  In MOND, it would be convenient for these baryons
to be associated with groups and clusters of galaxies.  I am not aware of a satisfactory idea for what form these dark baryons might take in either context.

Closer examination of Fig.~\ref{BTF} reveals that while groups and clusters are offset from the MOND prediction, they parallel it.
That is to say, the slope of the mass--circular velocity relation is well explained by MOND at all scales.  The different slope predicted by \LCDM\ is
only acceptable over a narrow dynamic range that happens to correspond to the scale of the richest clusters.  

Another test is provided by the collision velocities of sub-clusters.  
The high observed collision velocity of the components of the bullet cluster \cite{clowe} is not expected in \LCDM\ \cite{komatsubullet},
though it may result from hydrodynamical effects \cite{bullethydro}.
MOND naturally produces the high speed of the collision \cite{angmcg}.  Such high speed collisions are predicted to be more common than in \LCDM. 

The mass budget of rich clusters of galaxies works out better in \LCDM\ than in MOND.  On galaxy scales, the opposite is true.
Neither theory is completely satisfactory at the group scale.  Further tests may be afforded by measuring collision velocities of merging clusters,
but no firm statement can be made about this subject at present.  For these reasons, I consider the evidence on group and cluster scales to be ambiguous. 

\subsection{On Objectivity}

I have chosen to briefly review the evidence from groups and clusters here because these are often cited as providing the strongest evidence against MOND.
This evidence is not so clearly lopsided in favor of \LCDM\ as it is sometimes portrayed.
This has been my general experience.
While MOND certainly has legitimate problems, there are also many exaggerated claims to have falsified it.  

As discussed in \S \ref{sec:DMright} and \S \ref{sec:MONDright}, there are two distinct world-views that are mutually incompatible. 
It is extremely difficult to see both points of view.  This makes it challenging to evaluate the credibility of claims that confirm
or contradict what we expect to hear: we are all susceptible to cognitive dissonance, where we give greater weight to evidence supporting the position
we already hold.

If we are to remain objective, it is necessary to be willing to change one's mind.
Originally, I thought the solution to the mass discrepancy problem \textit{had} to be CDM.
Now I am not so sure.  Criteria that would again persuade me that CDM is indeed correct 
are (1) a convincing laboratory detection of appropriate dark matter particles, and (2) a viable explanation for the observed MONDian phenomenology.  

\section{Structure formation in MOND}

Considerable effort has been expended in developing the \LCDM\ structure formation paradigm.  A great amount of work is ongoing in trying to simulate the
formation of galaxies in this context.  In MOND, the successes at galaxy scales are well established.  Rather less effort has been exerted on structure formation,
and is less well known.  I review some of the successes and failures of MOND's predictions here.

First, one must recognize that MOND as a theory is an extension of Newtonian dynamics \cite{BM84}.  It is not a relativistic theory, and as such makes no
definitive predictions about cosmology.  Of course, if it is a correct aspect of Nature in the same sense as Newtonian dynamics, then ultimately an appropriate
relativistic theory encompassing both General Relativity and MOND must be found \cite[e.g., TeVeS:][]{TeVeS}.  This subject is beyond the scope of this
review \cite[see][]{LivRev}.

Nevertheless, it is well known that many results can be obtained conventionally by solving Newton's equations in an expanding background.  
Essentially all cosmological simulations are made in this way.  The same approach may be attempted with MOND.

MOND only pertains in the limit of low accelerations.  An obvious starting point is to assume that it is not active in the early, high density universe.
If so, all of the usual early universe results are retained.  In particular, BBN proceeds as usual (there is no reason to expect the expansion rate to be different
at early times), leading inevitably to the CMB (plasma physics is unaltered).  Structure presumably grows from the small initial perturbations observed in the 
CMB.  Whether these are provided by an earlier period of Inflation is irrelevant here.

\subsection{The formation of large scale structure}

The early history of a MOND universe must first diverge from conventional cosmology around the time of recombination ($z = 1088$).  
Since $\Omega_m \approx \Omega_b$ (modulo the neutrino mass) rather than $\Omega_m > \Omega_b$, the time of matter--radiation equality occurs 
after recombination rather than before.  In \LCDM, $z_{eq} \approx 3300$, but without dark matter this becomes $z_{eq} \approx 400$, depending
on the precise value of $\Omega_b$ \cite{MearlyU,SK01}.

\begin{figure}
\includegraphics[width=5.5in]{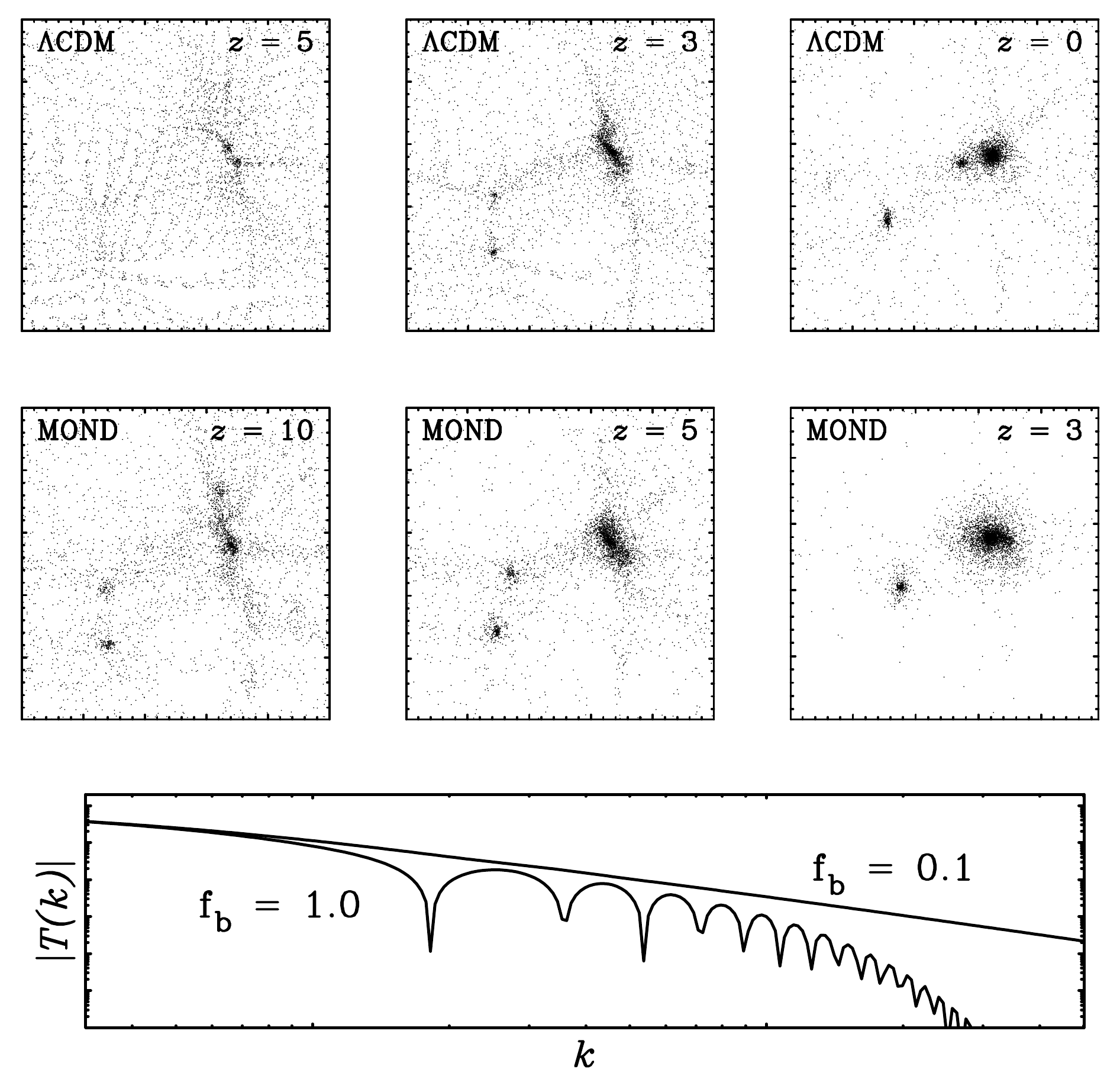}
\caption{The formation of large scale structure in \LCDM\ (top) and MOND (middle) according to the MLAPM N-body code \cite{KnebeGibson04}.
Similar structure forms in both cases, but starting from the same initial conditions develops more rapidly in MOND (note redshifts in each panel).
The bottom graph shows the transfer function emerging after recombination for both models: \LCDM\ with a baryon fraction of 10\% (smooth curve) 
and a purely baryonic model ($f_b =1$).  The presence or absence of sharp baryonic features provides a good test \cite{MearlyU,DodTVS},
but mode mixing during the non-linear growth of structure in MOND precludes application of this test at low redshift.  I expect structure to form
early in MOND, forming large galaxies early, providing the UV photons that reionized the universe \cite{M99,M04}, and establishing the cosmic
web early on.  The prediction of sharp baryonic features may be testable at high redshift, for example with observations of the 
redshifted 21 cm line \cite{21cmrecombination}.}
\label{LSS}
\end{figure}

Consequently, the photon field retains its grip on the baryons for a period after recombination, suppressing the initial formation of structure.
While in \LCDM\ the formation of the first small dark matter halos is already under way at the time of recombination, nothing can happen in MOND
until $z < z_{eq} \approx 400$.  

The late start to structure formation in MOND is exacerbated if neutrinos have substantial mass.  If $m_{\nu} \sim \sqrt{\Delta m_{\nu}^2}$, then neutrinos
do not have sufficient mass to affect the formation of structure.  If instead the neutrino mass is near its experimental upper limit of order $\sim 1$ eV,
then they are very important and act to further delay structure formation.  This is the reason that cosmological limits on the neutrino mass are
more restrictive than those from experiments in the conventional context:  even a modest neutrino mass is fatal to structure formation theory.  
However, these limits do not apply in MOND.  Indeed, a non-negligible neutrino mass might be turn out to be helpful in preventing MOND
from over-producing structure.

While the growth of density perturbations proceeds in the linear regime conventionally, this is not the case in MOND.
Once portions the universe begin to enter the low acceleration regime, they find themselves subject to the strongly non-linear MOND force law \cite{sanderscosmo}.
This obviates the use of perturbation theory.  After a slow start, the formation of structure in MOND is fast and furious.
Massive galaxies form at $z \approx 10$, the cosmic web is already distinct by $z \approx 4$, the first massive clusters assemble around $z \approx 2$, 
and large voids have been swept empty 
by $z = 0$ \cite{sanderscosmo,MearlyU,M99,M00,SK01,sandersPk,nusser,M04,KnebeGibson04,skordis06,llinares08,AD11,ADFvdH13,HarleyClusters}.

Fig.~\ref{LSS} illustrates the progression of structure formation in both \LCDM\ and MOND.  From the same set of initial fluctuations, very similar structures form.
This is not surprising, as MOND is basically just a tweak to the Newtonian force law.  The chief difference is the rate at which these structures form.  A comparable
degree of evolution occurs at higher redshift in MOND than in \LCDM.

The observer's experience is that structure has persistently been more developed at high redshift than anticipated.  Indeed, this was part of the reason to 
rehabilitate $\Lambda$, to give a little more time for structure to start forming in the early universe \cite{MoFukugita}.  This is as least qualitatively consistent
with the expectation in MOND.  The downside of the rapid growth of structure is the risk of overproducing structure by $z=0$.  The amplitude of fluctuations
$\sigma_8$ comes out about a factor of two too high in straightforward simulations \cite{nusser,KnebeGibson04}.  This is simultaneously problematic and impressively
close for a first attempt: it is only a factor of two off after non-linear growth by a factor of $10^5$.
SCDM was considerably further off at a similar stage of development \cite{karlfisher}.  

In any event, it is difficult to imagine that the growth rate of structure is the same in MOND and \LCDM, even if it is possible to arrange for them to equate
at both recombination and $z=0$.  A test is thus provided by the growth factor as a function of redshift \cite{BFS14}.  Informed by the limited numerical simulations
that have been done, I expect that structure forms earlier in MOND than in \LCDM.  In contradiction to this, it has been argued that one possible relativistic parent 
theory for MOND, TeVeS \cite{TeVeS}, would form structure later \cite{Dodelsonproblem}.  
Both of these statements assume that MOND is a modification of gravity, but it might instead be a modification of inertia \cite{milgrom83}.
Clearly there is a lot to sort out both theoretically and observationally.

\subsection{Baryonic oscillations}

One potential way to distinguish between \LCDM\ and MOND is in the power spectrum $P(k)$ of galaxies.  At $z=0$, this is quite consistent with \LCDM\ \cite{zehavi05}.
In MOND, it is clear that structure will form rapidly by a large factor, but the shape of the power spectrum is not well specified \cite{SMmond}.

Fig.~\ref{LSS} shows the transfer function that emerges from the CMB at high redshift \cite[as published by][]{MearlyU}.  
In \LCDM, the dark matter smooths the power spectrum.  Without CDM, strong baryonic oscillations are frozen into the plasma.
Indeed, Ref.~\cite{Dodelsonproblem} has pointed this out as a problem for MOND.  However, as already noted in \cite{MearlyU},
the strongly non-linear growth of structure in MOND can mix initially independent modes $k$, so it is not obvious that the initial oscillations
persist to low redshift.  The situation should be better at early times, before much mode mixing has had the opportunity to take place.
One might therefore hope to observe such oscillations in a large scale structure survey of high redshift galaxies or quasars, or in the 21 cm recombination
signal \cite{21cmrecombination}.

\subsection{Fluctuations in the cosmic microwave background}
\label{cmbfluctuations}

Perhaps the best evidence for the existence of non-baryonic CDM is the power spectrum of temperature  fluctuations in the CMB.
As mentioned earlier, fits obtain $\Omega_{\mathrm{CDM}} > 0$ at $44\sigma$ \cite{planckXVI}.  These fits assume General Relativity tells us everything
we need to know in order to make this computation.  Provided this assumption is true, then one does indeed need some form of non-baryonic dark matter.
In principle it need not be a WIMP, but could be a sterile neutrino \cite{angussterile}.  Here I am more concerned with the adequacy of the assumption.

\begin{figure}
\includegraphics[width=5.5in]{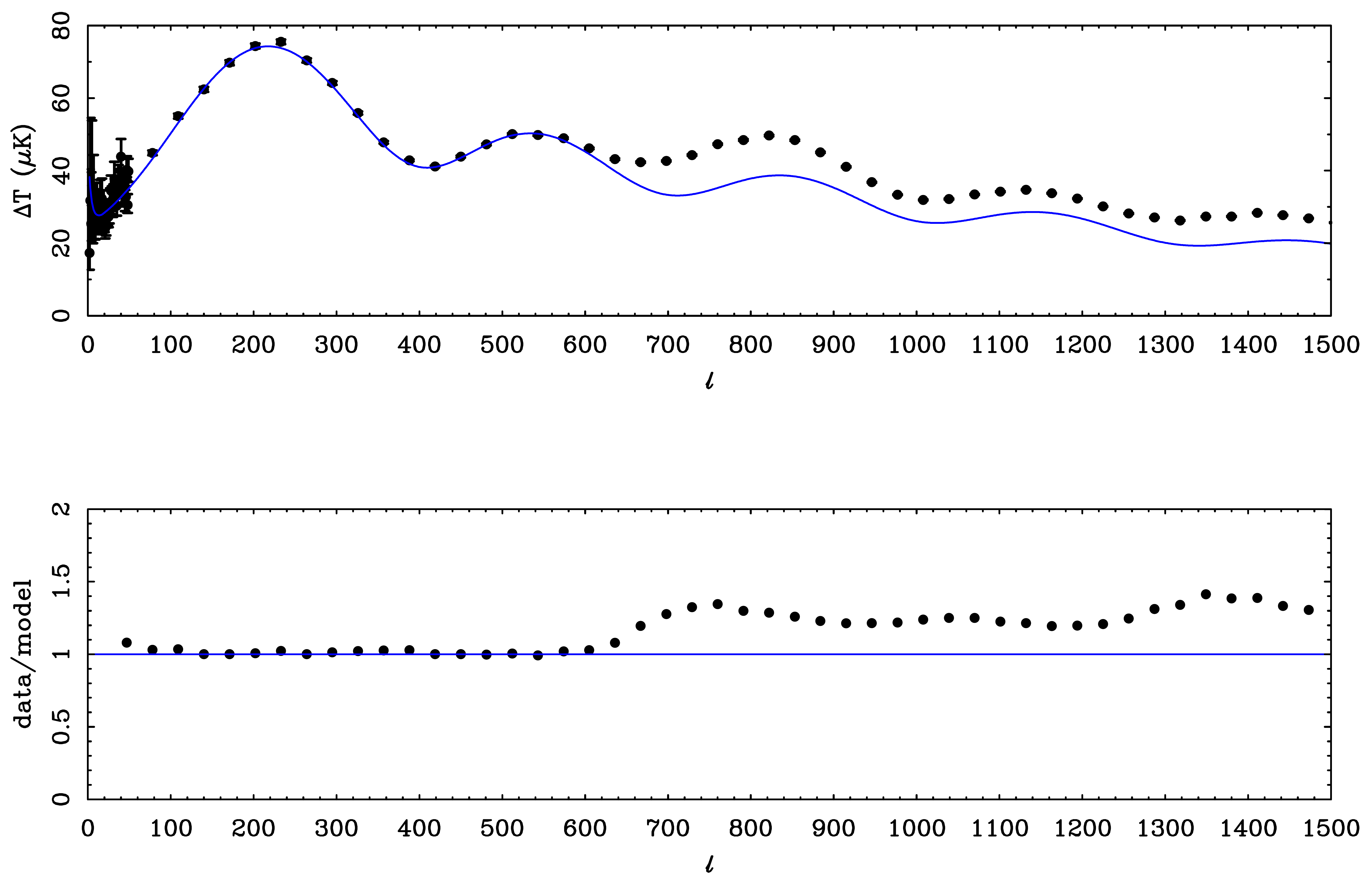}
\caption{The acoustic spectrum of temperature fluctuations in the cosmic microwave background (upper panel) as measured by Planck \cite[points]{planckXV} 
and the no CDM model of \cite{M04} (line).  The model is identically that published in 2004; it has not been modified in any way.  The only fit
parameter is a normalization in scale to match the position of the first peak. The shape of the spectrum was specified in 1999 by \cite{M99}.
This provided the only successful \textit{a priori} prediction of the amplitude ratio of the first to second peak \cite{M00}. 
This model is simply General Relativity without CDM. It is not a prediction of MOND per se, just the simplest possible \textit{ansatz} therefore.
It nevertheless provides an acceptable description of the data for $\ell < 600$.  For $\ell > 600$, the observed temperature fluctuations exceed
the simple model by $\sim 30\%$ (lower panel).  The true prediction of a relativistic MOND theory must necessarily deviate from the 
simple \textit{ansatz} \cite{M99}.  For example, early reionization was also predicted \textit{a priori}.  However, it is far from clear whether any
relativistic extension of MOND can explain the high amplitude of the third and subsequent peaks, which occurs naturally with CDM.}
\label{fig:CMB}
\end{figure}

Fig.~\ref{fig:CMB} shows the acoustic power spectrum from Planck \cite{planckXV} together with the prediction of a model devoid of CDM \cite{M99,M04}.
In the absence of CDM, baryonic damping dominates and one should see a spectrum in which each peak is smaller in amplitude than the one
preceding it.  When CDM is present, there is an additional forcing on the oscillations.  This manifests as the observed third peak exceeding the amplitude of the 
no-CDM model.

Modern CMB data, and particularly the third peak in the acoustic power spectrum, clearly falsify the simple no-CDM model.  Such a model was of course
never viable.  Nonetheless, it does serve as a useful starting point for what impact MOND might, and might not, have on the CMB.

The presence or absence of CDM is the most obvious, though certainly not the only, difference between \LCDM\ and MOND.
This simple fact was used by \cite{M99} to predict the minimum difference between the two.  This is surprisingly subtle.  The same
oscillations are expected in both theories, since these depend on plasma physics which is the same in either case.

\LCDM\ parameters were already known with some confidence by the late 1990s \cite{turner}.  Within these bounds, the most perceptible difference
at that time was in the expected amplitude ratio of the first-to-second peak.  Thanks to the net forcing term provided by CDM, the second peak was
predicted to be larger relative to the first with CDM than without \cite{M99}.  Indeed, without the flexibility introduced by CDM, the first-to-second peak
ratio was predicted to lie in a narrow range \cite{M99}: $2.2 < A_{1:2} < 2.4$.  With CDM, many peak ratios are possible, but for the parameters
known at that time, the expectation was $A_{1:2} < 1.9$.

The predictions above were made prior to the existence of any CMB data that constrained them.  These first appeared in 2000 \cite{boomerang}, and
were consistent with the no-CDM prediction \cite{M00}.  The observed value in more accurate data is $A_{1:2} = 2.34 \pm 0.09$ \cite{wmappeaks}.
This is consistent with the \textit{a priori} prediction of the no-CDM model.  

The first two peaks remain consistent with the 1999 prediction of the no-CDM model (Fig.~\ref{fig:CMB}).  Indeed, this much of the data can be fit with only
a single parameter:  the first peak position.  The model shown in Fig.~\ref{fig:CMB} is identically that from a decade ago \cite{M04}:
no adjustment has been made.  The shape of the power spectrum predicted by the model is fixed, so it is remarkable that the match to
even the first two peaks is this good.  

The \LCDM\ model did not predict the peak amplitude ratio \textit{a priori}, but does provide an excellent \textit{a posteriori} fit to the data \cite{planckXVI,WMAP7}.  
This is achieved by treating the baryon
density as a fit parameter that is permitted to vary outside the range previously allowed by BBN.  Prior to the existence of CMB data capable of constraining
the baryon density, no isotopic abundance suggested $\Omega_bh^2 > 0.02$ (Fig.~\ref{fig:BBN}).  In fits, the baryon density is driven to these higher values in order
to increase the damping enough to fit the second peak.  \LCDM\ thus `wins ugly' by having sufficient flexibility to accommodate the observations.  Despite
the many multipoles now observed, the fitting parameters available to the \LCDM\ model are quite sufficient to match any plausible CMB power spectrum.

The third and subsequent peaks are a clear victory for \LCDM.  The natural consequence of non-baryonic CDM is a net forcing term that
drives the oscillations away from the simple damping spectrum, as observed.  Indeed, the deviation sets in rather abruptly at $\ell > 600$,
almost as a step function (Fig.~\ref{fig:CMB}).  It is however not clear that this uniquely requires CDM, as it has been shown that a scalar field of the
sort invoked by TeVeS can have the same net driving effect on the oscillations as CDM \cite{skordis06}.  Perhaps this is a clue to the nature of a 
relativistic parent theory for MOND.

\subsection{Reionization \& CMB lensing}

It was pointed out from the beginning \cite{M99} that the no-CDM model must inevitably fail, even if MOND were correct.  The rapid growth of structure
expected in MOND has two late-time consequences for the CMB.  One is early reionization.  The other is an enhancement of gravitational lensing by structure
intervening between the surface of last scattering and an observer at $z=0$.

The signature of excess gravitational lensing is two-fold.  It creates an excess of power at low $\ell$ through the integrated Sachs-Wolfe effect. It also creates
an excess of power at high $\ell$.  Both of these effects have now been observed \cite{planckXV}.  This is qualitatively consistent with MOND.
Unfortunately, being more quantitative requires a relativistic theory.

The signature of early reionization is an enhanced optical depth to CMB photons due to Thompson scattering off free electrons.  
The optical depth is now well measured, implying a redshift of reionization $z_{reion} = 10.6 \pm 1.1$ \cite{WMAP7}.
This is considerably higher than was anticipated by \LCDM\ \cite[$z \approx 7$]{reionization}.  However, it was explicitly predicted that this should happen
in MOND \cite{M99,M04}.  The observed redshift of reionization corresponds nicely with the epoch of major galaxy formation \cite{hizgalform}
in MOND \cite{sanderscosmo}.

A potentially important detail is that reionization happens rather gradually.  
Though well underway at $z = 10$, it is not complete until $z \approx 6$ \cite{QSOreion}.
This may be an indication that the major source of ionizing photons is ordinary stars drawn from a normal stellar mass spectrum.
These can only do the job if given a running start, as in MOND.
In \LCDM, one needs to invoke supermassive Population III stars in order to make the few baryons that have formed into structures
by $z = 10$ about 50 times as efficient at producing ionizing photons as baryons are today.  
This tends to result in a rather abrupt epoch of reinoization \cite[see][and references therein]{M04}.
It should be noted that recent observations of extremely low metallicity stars \cite{FeHminus7} disfavor supermassive Population III stars,
which should produce a pattern of elemental abundances different from that which is observed.

In summary, modern CMB data clearly requires physics beyond a simple no-CDM model.  The amplitude of the third peak is naturally explained if that physics
is non-baryonic CDM.  On the other hand, the amplitude of the second peak and the epoch of reionization constitute successful \textit{a prioiri} predictions
in the context of MOND \cite{M99}.  Perhaps these are merely remarkable coincidences, or perhaps they are hints of a deeper theory.

\section{Summary}

There are at present two dialectically opposed explanations for the observed mass discrepancies in the universe.
In the concordance cosmology, \LCDM, the universe is filled with non-baryonic, dynamically cold dark matter.
In MOND, the mass discrepancy is ascribed instead to a change in the force law at a critical acceleration scale 
$a_{\dagger} \approx 10^{-10}\;\mathrm{m}\,\mathrm{s}^{-2}$.

I have reviewed some of the data pertinent to each paradigm.  
Where one makes clear predictions, the other tends to be mute.
This makes comparison of the two fraught. 
The conclusion one comes to depends on how one chooses to weigh the various lines of evidence.  

There are serious challenges for both ideas.
The acceleration scale $a_{\dagger}$ is clearly written in the dynamical data.
This is not natural to the scale free nature of CDM, so one important issue is whether plausible \LCDM\ models can be constructed to
accommodate this fact.  Here `plausible' is key.  There are many moving parts in galaxy formation models.  There is no doubt that they have
sufficient flexibility to fit any given set of data.  Since such models can do an enormous variety of things, yet the data for individual galaxies
do the one specific and unique thing predicted by MOND, fine-tuning galaxy formation models inevitably violates Occam's razor \cite{myrutgers,MdB98a}. 

The looming challenge for MOND is to find a satisfactory relativistic theory that reproduces all the successes of 
General Relativity as well as MOND in the appropriate limits.  This is no small task.  At present, there are many theories under consideration.
Whether any are satisfactory is too soon to judge.  However, it seems to me that \textit{if} MOND is true, then we are missing something
conceptual at a fundamental level.  Perhaps this is related to Mach's Principle and the origin of inertial mass, but this is simply speculation.

Finally, a fundamental question is whether non-baryonic CDM actually exists.  \textit{If} the concordance cosmology is correct, it must.
Contrawise, the non-existence of CDM falsifies the concordance cosmology.  

The situation is somewhat reminiscent of that of \ae ther in the nineteenth century.
Given what we know of cosmology today, non-baryonic dark matter must exist.
But does it?  We know there must be new physics, but of what kind?

The existence of the \ae ther was at least falsifiable.  It is not obvious that CDM meets this standard, and we teeter on the brink of the definition of science.
The existence of CDM is confirmable:  a clear laboratory detection of appropriate WIMPs would suffice.
However, the existence of dark matter is not falsifiable \cite[though see][]{MdB98a,kroupafalsify,KPM12}.  If we fail to find WIMPs, maybe it is axions.  
If not axions, we are free to invent another form of dark matter --- and another, and another, and so on, \textit{ad infinitum}.  CDM was invented for very good reasons.  
But if this hypothesis happens to be wrong, how do we tell?
 
\section*{Acknowledgements}

I thank Benoit Famaey, Pavel Kroupa, Federico Lelli, Moti Milgrom, Marcel Pawlowski, and Bob Sanders for helpful conversations.
I acknowledge Kev Abazajian and Frank van den Bosch for providing the impetus to revisit these issues.



\end{document}